\documentclass{phbauth}
\usepackage{graphicx}

\begin{document}

\begin{frontmatter}

\title{Ordered phases in spin-Peierls systems}

\author{G\"otz S. Uhrig}

\address{Institut f\"ur Theoretische Physik,
Z\"ulpicher Stra\ss{}e 77, Universit\"at zu K\"oln,
D-50937 K\"oln, Germany}

\begin{abstract}
The microscopic description of spin-Peierls substances is discussed.
Particular attention is paid to the ordered (dimerised and incommensurably
modulated) phases. Important points are the adiabatic and the antiadiabatic
 approach, generic soliton forms,
 elastic and magnetic interchain couplings. The wealth and the accuracy of
experimental information collected for the first 
inorganic spin-Peierls substance CuGeO$_3$ 
motivates this work.
\end{abstract}

\end{frontmatter}

The coupling of 1D magnetic and 3D
elastic degrees of freedom in spin-Peierls systems leads to
magnetically driven lattice distortions at low $T$ 
(see e.g.~\cite{bray83,bouch96}).
Without magnetic field the spin chains dimerise (D phase). 
For sufficiently large
magnetic field or by defects the homogeneous dimerisation
is modulated. Solitons occur: zeros of the dimerisation at
which $S=1/2$ spinons are localised \cite{uhrig99b,uhrig99c}.
The high-magnetic field, incommensurate (I) phase is characterised
by an array of these solitons.

Since we deal with coupled magnetic and elastic degrees
of freedom the microscopic Hamiltonian 
is $H= H_{\rm S}+H_{\rm SB}+H_{\rm B}$. The purely magnetic part is given by
\begin{equation}
H_{\rm S} = J\sum_{i,j} \left( {\bf S}_{i,j}{\bf S}_{i+1,j} + 
\alpha {\bf S}_{i,j}{\bf S}_{i+2,j}\right)
\end{equation}
where $i$ is the site index within a given chain $j$. The frustrating
next nearest neighbour coupling $\alpha J$ is included since
there is evidence that this term is important in CuGeO$_3$ 
\cite{riera95,casti95,fabri98a}.
The phonons are treated as free bosons
\begin{equation}
H_{\rm B} = \sum_{{\bf q}} \omega({\bf q}) b^\dagger_{{\bf q}} 
b^{\phantom \dagger}_{{\bf q}}\ .
\end{equation}
The Peierls phenomenon stems from the interaction $H_{\rm SB}$. Since the atomic
displacements are small we assume that they influence the magnetic exchange 
in linear order
\begin{eqnarray}
\label{kopplung}
H_{\rm SB} &=& \sum_{{\bf q}} 
A_{{\bf q}} (b^+_{{\bf q}}+b^{\phantom +}_{-{\bf q}})  \\
\mbox{with} \quad A_{{\bf q}}  &=& \sum_{{\bf k}} g({\bf q},{\bf k}) 
{\bf S}_{{\bf k}}{\bf S}_{-{\bf k}-{\bf q}} \ .
\label{g-def}
\end{eqnarray} 
Unfortunately, this Hamiltonian  is too complicated. 
So one argues that the 1D
magnetic subsystem strongly favours a distortion at wave vector $\pi$ 
which is underlined by the dimerisation below the transition
temperature $T_{\rm SP}$
and by the diverging dimer-dimer correlation at $\pi$ for $T\to 0$.
Hence those  phonons matter most which are in the vicinity of the actual lattice distortion
at ${\bf q}^{\rm d}$ for $T<T_{\rm SP}$. It
appears permissible to take their energy $\omega({\bf q}^{\rm d})$
as the energy of {\it all} the phonons neglecting the phonon dispersion.
This does {\it not} imply that the 3D character is lost since in 
generic lattices with basis even the local phonons are extended objects
(for the polarisation vectors of CuGeO$_3$ see 
\cite{brade96a,brade98a,werne99}).
This fact leads to the important elastic coupling {\it between} adjacent
chains so that the spin-Peierls transition can take place at $T>0$.
A purely 1D model would not display a phase transition at finite $T$.
By the above simplifications we arrive at 
$H_{\rm B}= \omega\sum_{i,j}b^+_{i,j}b^{\phantom +}_{i,j}$.
The spin-phonon coupling in real space reads
$H_{\rm SB}= g\sum_{i,j} A_{i,j}(b^+_{i,j}+b^{\phantom +}_{i,j})$.
The operators $A_{i,j}$ contain spin operators not too far
from the point indexed by $i,j$, e.g.
$A_{i,j}=g ({\bf S}_{i,j}{\bf S}_{i+1,j}- {\bf S}_{i,j}{\bf S}_{i-1,j}+
f\sum_\pm({\bf S}_{i+1,j\pm1}{\bf S}_{i,j\pm1}-
{\bf S}_{i,j\pm1}{\bf S}_{i-1,j\pm1}))$ \cite{uhrig98b}.
 The parameter $f\, (|f|<1)$ measures how much
a phonon on chain $j$ influences also its adjacent chains $j\pm1$.

Now two routes can be followed. 
The first, more traditional, route consists in treating 
the phonons adiabatically, i.e.\ the distortions are considered as static
\begin{equation}
\label{static}
b_{i,j} \to \langle b_{i,j}\rangle=:\delta_{i,j}J/(2g)
\end{equation}
in the ordered phases.
Correspondingly, there is no influence of $H_{\rm SB}$
 {\it above}  $T_{\rm SP}$, see e.g.~Ref.~\cite{klump99a}.
The phonon propagators are renormalised in RPA
by the magnetic response in the dimer-dimer channel \cite{cross79}.
The transition is signalled by the vanishing of the renormalised
phonon frequency, the so-called phonon softening. This description
applies if the phonons are the {\it slow} degrees of freedom \cite{cross79}
which are renormalised by the {\it fast} magnetic 
degrees of freedom. An obvious condition
for the validity of this  scenario is $\omega < J$.

It was
shown by numeric investigation that the more restrictive condition
$\omega < \Delta$ applies where $\Delta$ is the gap in the D
phase \cite{caron96}. 
The gap $\Delta$ is a measure for the effect of the spin-phonon coupling, i.e.
$\Delta$ indicates up to which energy the magnetic  and
the phononic degrees of freedom are decisively altered by the coupling.
Hence the phonon cannot become soft if $\omega > \Delta$ holds. This view,
suggested in particular for the  spin-Peierls substance CuGeO$_3$ 
\cite{uhrig98b}, is corroborated by field theoretic results \cite{gros98}.

The second route consists in the antiadiabatic approach which considers
the spin system as {\it slow} and the phononic system as {\it fast}.
Then it is reasonable to eliminate the phonons in favour of an effective
spin model. Many works took actually this approach 
\cite{pytte74a,weiss99b} in leading order
in $J/\omega$. Couplings dependent on $T$ arise in the next-leading
order \cite{uhrig98b}. The transition into the ordered phase does
not occur due to the softening of a phonon but due to the tendency of the
effective spin model towards dimerisation. Considering a
{\it single} chain phonon-induced frustration above its critical
value $\alpha_{\rm eff} > \alpha_{\rm c}\approx 0.241167(5)$ \cite{egger96}
 drives the  dimerisation \cite{uhrig98b,bursi99a,weiss99b}. 
Hence a finite  spin-phonon
interaction is necessary to achieve the spontaneous symmetry breaking
\cite{augie98b,sandv99a,kuhne99}.

In  Sect.~1 we compare the adiabatic and the antiadiabatic
route to the spin-Peierls transition. In Sect.~2
the I phase will be discussed in detail. A summary concludes this article.

\section{Mean-field, nonadiabaticity and phasons}
Most adiabatic approaches use the Hamiltonian
\begin{equation}
\label{ham-mf}
H = J\sum_i\left[(1+\delta_i){\bf S}_{i+1}{\bf S}_{i} +
  \alpha {\bf S}_{i+2}{\bf S}_{i} + \frac{K}{2} \delta_i^2
\right]
\end{equation}
with some spring constant $K$ (see e.g. \cite{feigu97,schon98}); the chain
index has been dropped. The elastic interchain coupling does
 not appear since its main effect can be
absorbed in the 1D model as long as the modulations on neighbouring chains
occur in phase. An important exception to this rule are solitons induced
by impurities \cite{hanse98,augie98c,uhrig99c}.
The local distortions $\delta_i$ in Eq.~(\ref{ham-mf}) 
 are subject to the constraint $\sum_i\delta_i =0$. The ground state
energy $\langle H\rangle$ is minimised with respect to the $\delta_i$
leading to
\begin{equation}
\label{delta-cal}
K\delta_i = -\langle {\bf S}_{i+1}{\bf S}_{i}\rangle +
\overline{\langle {\bf S}_{i+1}{\bf S}_{i}\rangle}
\end{equation}
where the overline stands for the sample average to ensure the constraint.

\begin{figure}[htb]
\begin{center}\leavevmode
\includegraphics[width=0.8\linewidth]{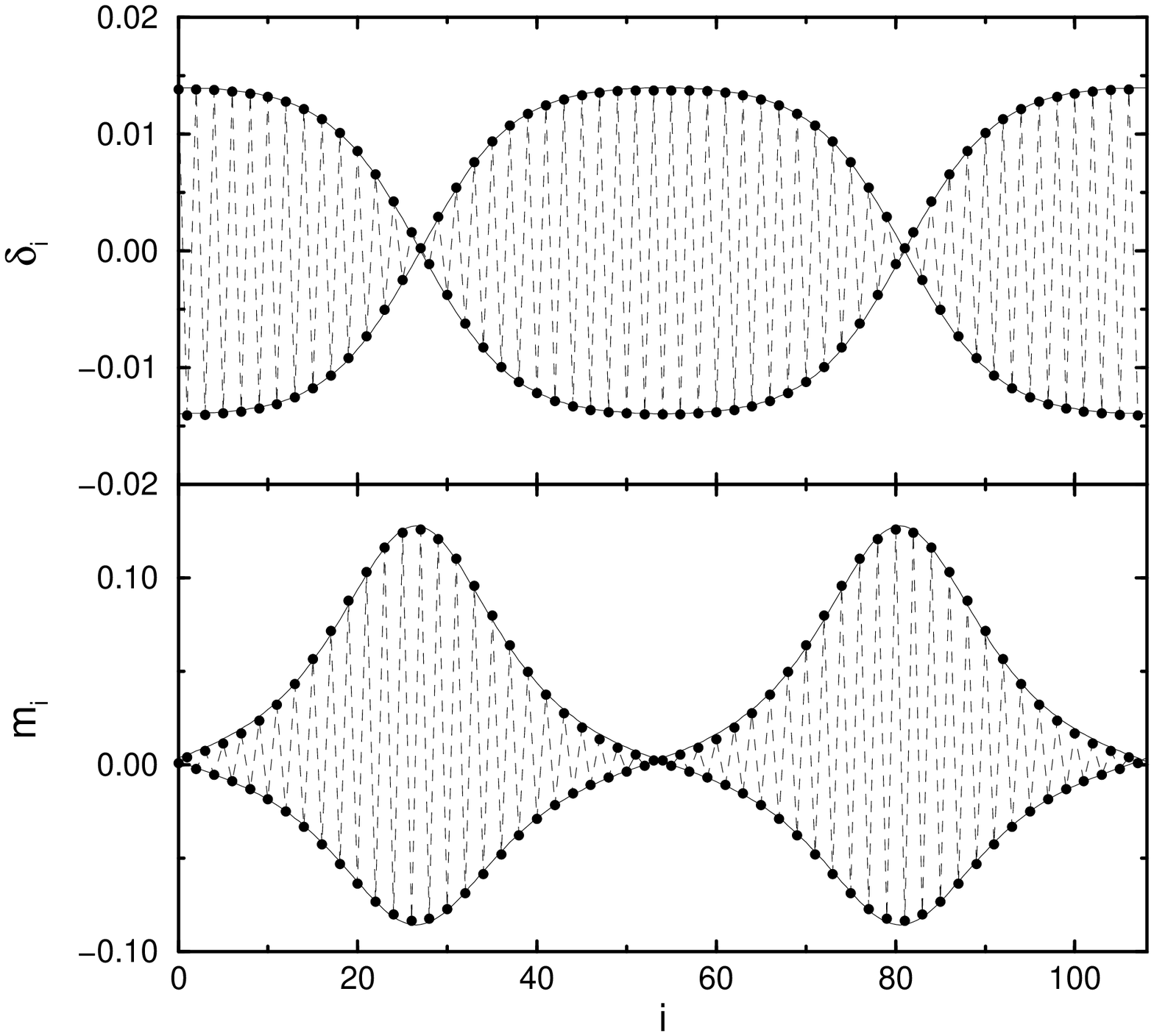}
\caption{Upper panel: local distortions. Symbols: 
 DMRG-result at $K=18J$, $\alpha=0.35$;
 solid line  from
 (\protect\ref{fit1b}) with $\delta= 0.014$, $k_{\rm d}=0.959$,
 $\xi_{\rm d} = 10.5$.
Lower panel: local magnetisations; solid line from
 (\protect\ref{fit1a}) with $W=0.21 $, $R=5.0$, $k_{\rm m}=0.992$, 
$\xi_{\rm m} = 7.9$.}
\label{fig-soliton}
\end{center}
\end{figure}

For $S_{\rm tot}=0$, i.e.~in the
D phase, the solution to the minimisation of Eq.~(\ref{ham-mf})
reads $\delta_i=(-1)^i\delta_0$ where the amplitude $\delta_0$
is proportional to the expectation value of the dimerisation operator
$H_{\rm D}=\sum_i(-1)^i{\bf S}_{i+1}{\bf S}_{i}$, 
i.e.~$\delta_0=\langle H_{\rm D}\rangle/K$ (see e.g.~\cite{knett99a}).
For finite magnetisation solitons occur. A generic solution for $S_z=1$
is displayed in Fig.~\ref{fig-soliton} \cite{uhrig99b}.

What does one obtain in the antiadiabatic limit?
For simplicity we assume $A_{i,j}=g(:{\bf S}_{i+1,j}{\bf S}_{i,j}:
+f\sum_\pm :{\bf S}_{i+1,j\pm1}{\bf S}_{i,j\pm1}:)$ where the colons
stand for the operator reduced by its expectation value
 ($:X: = X-\langle X\rangle$). This is to
prevent ``trivial'' renormalisation of the bare coupling constants.
Applying a unitary transformation eliminating $H_{\rm SB}$
one obtains in the two leading orders 
additional terms in the spin Hamiltonian 
$\Delta H_X={\rm O}(g^2/\omega)$
and $\Delta H_Y={\rm O}(g^2J/\omega^2)$ \cite{uhrig98b}
\begin{eqnarray}
\Delta H_X &=& \frac{-1}{\omega}\sum_{i,j}  A^+_{i,j}A^{\phantom +}_{i,j}\\
\Delta H_Y &=& 
\frac{1}{2\omega^2}\coth\left(\frac{\omega}{2T}\right) \sum_{i,j} 
\left[ A^+_{i,j}, {\rm L}A^{\phantom +}_{i,j}\right]\ ,
\end{eqnarray}
where ${\rm L}$ is the Liouville operator, i.e. the commutation
with $H_{\rm S}$. In $\Delta H_Y$ the interchain terms in $A_{i,j}$ do not
generate qualitatively new terms. In $\Delta H_X$, however, the  interchain terms
lead to terms 
\begin{equation}
\label{inter}
H_{\rm int} = -\frac{2g^2f}{\omega}\!\sum_{i,j}\!:\!{\bf S}_{i+1,j}{\bf S}_{i,j}\!:
:\!{\bf S}_{i+1,j+1}{\bf S}_{i,j+1}\!:
\end{equation}
coupling adjacent chains. It is obvious that via $H_{\rm int}$
the dimerisation on one chain favours dimerisation on the adjacent
chain. So in the antiadiabatic description the part (\ref{inter})
is responsible for the phase transition at $T>0$ since
it creates the coherence of the dimerisation pattern throughout the
whole sample.

It is often argued that the adiabatic phonon description works very
well, see e.g.~\cite{klump99a,werne99,gros98}. The pieces of evidence,
Ginzburg criterion, susceptibility and entropy fits, do indeed
favour a mean-field description. They do not, however, require
 an adiabatic phonon description. In view of  $\omega \gg \Delta$ in CuGeO$_3$ 
\cite{uhrig98b},
we advocate the {\it antiadiabatic} phonon description. 
This is combined with a chain-mean field treatment (CMF) since 
$T_{\rm SP}/J \approx 0.1$ is small.

 Within 
each chain an effective Hamiltonian as derived previously 
\cite{uhrig98b} is used. Between the chains
the contribution (\ref{inter}) is treated in
mean-field. Let us abbreviate
\begin{equation}
\label{delta-def}
\delta_i :=  -\frac{2g^2f}{\omega J}\sum_{\pm}
\langle :{\bf S}_{i+1,j\pm1}{\bf S}_{i,j\pm1}: \rangle \ .
\end{equation}
Then the $\delta_i$ can be interpreted as local distortions 
as in the adiabatic description, cf.~(\ref{ham-mf}). Defining
\begin{equation}
\label{K-def}
K_{\rm eff} := \left|\frac{\omega J}{4g^2f} \right|\ .
\end{equation}
the equivalence is pushed one step further.
Inserting Eq.~(\ref{K-def}) in Eq.~(\ref{delta-def}) and 
treating both neighouring chains as equal yields in CMF the {\it same}
equation (\ref{delta-cal}) as in the 
adiabatic treatment. This is true in the D phase and in
 the I phase as long as the modulation stays close
to $\pi$.

The ``good'' equivalence between the adiabatic treatment and 
the antiadiabatic treatment with chain-mean-field provides
 an explanation for the success of the former in 
spite of the high phonon energies. But it is 
{\em not} the adiabatic treatment in its own right which works
so well. There are two major differences between the adiabatic
treatment and the seemingly equivalent antiadiabatic treatment.
(i) The in-chain couplings are partly phonon-induced
and hence they depend on $T$ \cite{uhrig98b,fabri98b}. (ii)
In the adiabatic treatment $K$ is related to {\it in-chain} properties: 
$K= \omega J /(2g^2)$ (obtained  inserting Eq.~(\ref{static})
in $H_{\rm SB}$ and $H_{\rm B}$). The constant $K_{\rm eff}$ in 
Eq.~(\ref{K-def}) refers to an {\it inter-chain} property since
the spatial extent of the phonons as measured by $f$ enters decisively.

What are the points in favour of the antiadiabatic approach for CuGeO$_3$?
Besides the  relation between $\Delta\approx25$K and $\omega\approx150$K
\cite{uhrig98b} and the $T$-dependent couplings \cite{fabri98b} there
is the observation that the zero-point motion of lattice exceeds
the static distortions considerably \cite{werne99}. In particular in
the I phase, this zero-point 
motion has an important effect on the alternating local magnetisation 
\cite{uhrig98a,uhrig99b}.
 On the basis of results as in Fig.~\ref{fig-soliton}
NMR peaks can be understood \cite{fagot96}. In adiabatic calculations
the amplitudes of the alternating component, however, is found to be
much larger (factor 4 to 6) than in experiment. Theory and experiment can be
reconciled if the zero-point motion of the ``phonons'' of the soliton
lattice, the so-called phasons, are taken into account. This motion leads
to a reduction of the alternating component \cite{uhrig99b} which 
can be mimicked by an average of the magnetisation of adjacent sites
\cite{uhrig98a}. The phasons are connected to the spontaneous breaking
of the quasi-continuous symmetry of translating the modulations along
the chains. Such translations require that the distortions are also
dynamic. So the phasonic zero-point motion
 provides strong evidence for the nonadiabatic character
of the ordered phases in CuGeO$_3$.

\section{Incommensurate phase}
Turning to the solitons in the I phase
we first state the standard theory  \cite{nakan80}.
The original spin Hamiltonian (\ref{ham-mf})
is mapped by a Jordan-Wigner transformation \cite{jorda28}
 to a fermionic model
which in turn can be reduced to an effective 
bosonic continuum model at low energies
 \cite{nakan80,delft98}
\begin{eqnarray}
\nonumber
H &=& \frac{v_{\rm S}}{2\pi}\int \Big(
K(\pi\Pi)^2 + \frac{1}{K} (\partial_x\phi)^2
\Big) \mathrm{d}x\\ \label{boson}
&&+\int\Big(-\delta(x) \cos(2\phi)+\frac{K}{2} \delta^2(x)
\Big) \mathrm{d}x
\ ,\quad
\end{eqnarray}
where $\delta(x)$ is  the alternating component
of the distortion. Nakano and Fukuyama used the self-consistent harmonic 
approximation $\hat \phi \to \phi_{\rm cl} + \hat 
 \phi_{\rm fl}$ distinguishing a classical number $\phi_{\rm cl}$
and a fluctuating operator  part $\hat \phi_{\rm fl}$
with $\langle \hat \phi_{\rm fl}\rangle =0$. 
Treating the fluctuation in harmonic approximation leads  
to $\cos(2\hat\phi) \to \mathrm{e}^{-2\sigma}\cos(2\phi_{\rm cl}) 
(1-2\hat\phi_{\rm fl}^2)$ with 
$\sigma:= \langle\hat\phi_{\rm fl}^2(x) \rangle$ motivated by 
the WKB approach by Dashen {\it et al.} 
\cite{dashe74a}.
Varying the ground state expectation value of
(\ref{boson}) with
respect to $\delta(x)$ and to $\phi_{\rm cl}$, respectively, yields
\begin{eqnarray}
0 &=& -\cos(2\phi_{\rm cl})\mathrm{e}^{-2\sigma} + K\delta \\
0 &=& v_{\rm S}/(\pi K)\partial^2_x \phi +2\delta\sin(2\phi_{\rm cl})
\mathrm{e}^{-2\sigma}\ ,
\end{eqnarray}
which leads finally to
\begin{equation}
\label{nonlinear}
2\partial^2_y \phi_{\rm cl} = \sin(2\phi_{\rm cl})\cos(2\phi_{\rm cl})\ ,
\end{equation}
where $y=x/\xi$ for the correlation length $\xi$.
The {\em spatial} dependence of $\sigma$ is neglected
 \cite{nakan80}. Eq.~(\ref{nonlinear})
has the soliton array solutions \cite{zang97,uhrig99b}
(${\rm sn}, {\rm cn}, {\rm dn}$: elliptic Jacobi functions)
\begin{eqnarray}
\nonumber
m_i &=& W\{  
{\rm dn}(r_i/(k_{\rm m}\xi_{\rm m}),k_{\rm m})/R\\
\label{fit1a}
&& 
+(-1)^i {\rm cn}(r_i/(k_{\rm m}\xi_{\rm m}),k_{\rm m})\}/2 \\
\delta_i &=& (-1)^i \delta\ 
{\rm sn}(r_i/(k_{\rm d}\xi_{\rm d}),k_{\rm d})\ ,
\label{fit1b}
\end{eqnarray}
with $\xi_{\rm m}  =  \xi_{\rm d}$ and $k_{\rm m} = k_{\rm d}$.

In the  fits in Fig.~\ref{fig-soliton} one observes 
that $\xi_{\rm m}  =  \xi_{\rm d}$
does {\em not} hold. In addition,
there is no good reason to neglect the spatial dependence
of $\sigma$ (cf.~Ref.~\cite{fabri98c}).
Considering it leads to an extra factor $\mathrm{e}^{-4\Delta\sigma}$
on the right side of (\ref{nonlinear}) where $\Delta\sigma$ is the difference
to the fluctuations in the ground state. Previous results \cite{uhrig99c},
(not yet self-consistent: fluctuations on top of the
Nakano/Fukuyama solution), showed that the spatial dependence of 
$\Delta\sigma$ is indeed important. The
conjecture that the numerical finding $\xi_{\rm m}\neq\xi_{\rm d}$
\cite{uhrig99b} is related to the spatial dependence of 
$\Delta\sigma$ could be corroborated.

\begin{figure}[htb]
\begin{center}\leavevmode
\includegraphics[width=0.8\linewidth]{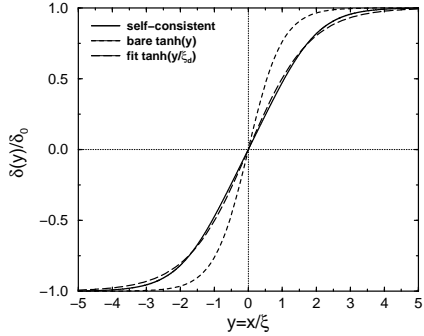}
\caption{Distortion of a single soliton
with (solid line), without (dotted line) spatial dependence of
$\Delta\sigma(x)$; fit with 
relativ width $\xi_{\rm d}=1.80$}
\label{fig-verzerr}
\end{center}
\end{figure}

\begin{figure}[htb]
\begin{center}\leavevmode
\includegraphics[width=0.8\linewidth]{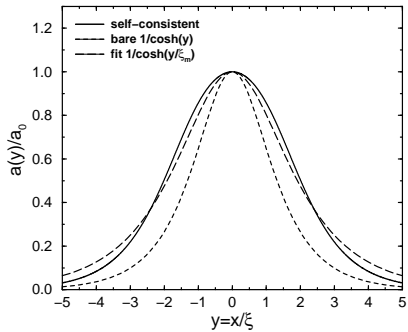}
\caption{Alternating magnetisation
 of a single soliton
with (solid line), without (dotted line) 
 spatial dependence of $\Delta\sigma(x)$; 
 fit with relativ width $\xi_{\rm m}=1.45$}
\label{fig-altern}
\end{center}
\end{figure}

Figs.~\ref{fig-verzerr} and \ref{fig-altern} show the results of a
fully self-consistent calculation.
Clearly the spatial dependence of the fluctuations makes the kink
smoother. This was expected since the fluctuations are maximum in the
center of the soliton (see Fig.~4 in Ref.~\cite{uhrig99c}).
Most interesting is the fact that the distortion is smoothened more
than the alternating magnetisation. From the fits one finds the ratio
$\xi_{\rm d}/\xi_{\rm m} =1.24$ in very good agreement
with the numerical result at critical frustration $\alpha_c$
(see Fig.~5 in Ref.~\cite{uhrig99b}). Hence the discrepancy between
the existing continuum theory of spin-Peierls solitons and the numerical
results is removed when the spatial dependence of the fluctuations are
considered properly.

\begin{figure}[htb]
\begin{center}\leavevmode
\includegraphics[width=0.8\linewidth]{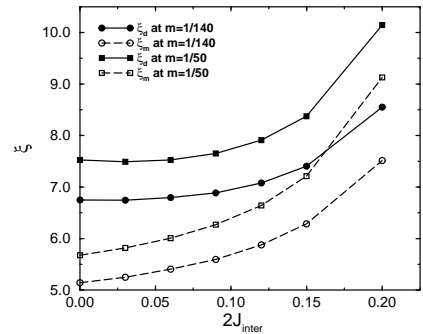}
\caption{Distortive ($\xi_{\rm d}$) and magnetic ($\xi_{\rm m}$)
soliton widths at two average magnetisations $m$ at $\alpha=0.35,K=10.2$.}
\label{fig-jinter}
\end{center}
\end{figure}

In spite of the progress just shown the agreement to experiment
is not yet perfect. The soliton widths observed experimentally still
exceed the theoretical ones \cite{uhrig99b}. As a next step towards
a quantitative description we include {\em magnetic} interchain couplings
as they are present and non-negligible in CuGeO$_3$ \cite{uhrig97a}.
They are treated in mean-field approximation so that only the
local magnetisations matter. Apart from this extension the Hamiltonian
(\ref{ham-mf}) is minimised since the antiadiabatic limit
leads to the same equations. In Fig.~\ref{fig-jinter}
the dependence of the soliton widths on an antiferromagnetic 
coupling $J_{\rm inter}$ between adjacent chains in a plane is shown.
Clearly, all widths are enhanced. The additional coupling
strongly favours the alternating magnetisation close to the center of the 
soliton. Thus the 2D coupling induces an increase of this region thereby
enlarging the soliton \cite{zang97}. So the agreement with experiment
will be improved by taking the 2D character of the magnetic couplings
into account. Yet further questions still remain as the ratio between
distortive and magnetic soliton width is decreased on increasing 
$J_{\rm inter}$ in contrast to experiment.

\section{Summary}
In this work we contrasted the adiabatic approach to
spin-Peierls transitions with the antiadiabatic approach.
If the latter is complemented by a chain-mean-field treatment
both approaches yield very similar equations. But the interpretation
of the parameters differ and so do certain physical phenomena such as
$T$-dependent couplings and local magnetisations. For CuGeO$_3$
the antiadiabatic approach was advocated.

In the incommensurate phase the form and the amplitudes
of solitons were discussed.
The extension of the existing
continuum theories as required by the numeric results was presented.
Further steps towards a quantitative description of the I phase
of CuGeO$_3$ were shown.

\begin{ack}
I like to thank C. Berthier, J.P. Boucher, B. B\"uchner, A. Dobry, 
A. Kl\"umper, T. Lorenz, U. L\"ow, M. Horvati\'c,  Th. Nattermann
for fruitful discussions, B. Mari\'c and F. Sch\"onfeld for  DMRG 
data and E. M\"uller-Hartmann for generous support.
Financial support of the DFG by the SFB 341 and by SSP 1073 is acknowledged. 
\end{ack}

\end{document}